\documentclass{icrc29}
\usepackage{graphicx,amssymb,amsmath,times}
\begin{document}

\title{
\bf {Measurement of the Sky Photon Background Flux at the Auger Observatory}}

\author[R. Caruso et al.] {R. Caruso, A. Insolia, S. Petrera, P. Privitera, F. Salamida and  V. Verzi
\newauthor for the Pierre Auger Collaboration}
    
%\correspondence{S.Petrera (sergio.petrera@aquila.infn.it)}
\presenter{Presenter: S.Petrera (sergio.petrera@aquila.infn.it), \  
ita-petrera-S-abs2-he15-poster}

\maketitle

\begin{abstract}
The sky photon background flux has been measured at the southern Auger site in Malarg\"ue, 
Argentina, using the observatory's fluorescence detectors (FD).  The analysis is based on ADC variances of 
pixels not triggered by the First Level Trigger. Photon fluxes are calculated for each 
individual pixel at each telescope. The statistics from each night of data taking allows 
a study of local variations in the  photon flux. 
Results show a clear dependence of the flux on elevation angle. Time variations, possibly 
related to different atmospheric conditions, do not mask this dependence. In particular 
the flux excess above the horizon shows a rather stable and reproducible behaviour with 
elevation. 
Correlation of this dependence with atmospheric parameters can be of interest as it offers
the promise of extracting those parameters directly from FD data, 
thus allowing cross checks with independent methods based on different monitoring devices. 
\end{abstract}

\section{Introduction}
\label{sec:intro}
 The Fluorescence Detector (FD) of the Pierre Auger Observatory \cite{EA}
is composed of a set of 24 telescopes, each measuring the 
longitudinal development of cosmic ray showers in the atmosphere above 
ground. The telescopes are arranged in four perimeter buildings (Eyes), each housing 6 telescopes, which overlook the Surface Detector (SD) array. 
In the focal surface of each telescope mirror, a camera 
detects the light on an array of 20$\times$22 photomultiplier pixels. Each pixel covers a 1.5$^\circ$ diameter hexagonal portion of the sky.
The first two buildings at Los Leones and Coihueco have been completed, with their 12 FD telescopes now taking data.  

The sky photon background flux has been measured using data taken during FD runs. 
The standard data acquisition records signals from
all triggered pixels and a fraction of pixels not triggered. This analysis is based on ADC 
variances of 
pixels not triggered by the First Level Trigger. Sky background photons are the main
source of FD noise random triggers. Therefore the study of their flux is of interest for
both monitoring the behaviour of the detector, and correlating the sky brightness on a 
pixel-to-pixel basis with atmospheric conditions.

\section{The method}
\label{sec:meth}

Using the theory of photon statistics,
the sky photon background flux can be directly derived from ADC variances \cite{HG}. 
The adopted analysis steps are the following:

\noindent 1. {\it Run selection:} 
Only moonless clear nights have been used. 
The data taking period extends from August through December 2004.
Short runs (of duration less than 1 hour)
have been excluded, since run stops have been usually related to hardware or acquisition 
problems.

\noindent 2. {\it Calculation of the sky background variance in ADC counts:} 
The ADC variance is the sum of sky background variance and electronic noise variance:
$$ [\sigma^2_{ADC}]^{bckg} = [\sigma^2_{ADC}]^{sky} + [\sigma^2_{ADC}]^{ele} \qquad .$$
In order to estimate electronic noise, we used the background files acquired with closed 
shutters every night before data acquisition. The sky background variance is then derived 
by subtracting from each pixel the relevant noise variance.

\noindent 3. {\it Conversion of sky background variance from ADC counts to photoelectrons:}
$$ \sigma^2_{phe} = [\sigma^2_{ADC}]^{sky} / A_G^2 $$
where $A_G$ is the absolute gain ($ADC~counts$ per photoelectron). 
Individual pixel are calibrated using an end-to-end procedure exploited with
a device known as the {\it drum calibrator} \cite{jeff}. This method provides a conversion
between  ADC counts and photons at the telescope diaphragm for each pixel. In terms of these calibration
constants $C_{FD}$, the absolute gain is: 
$$A_G = \frac{1}{C_{FD} \cdot f \cdot Q} $$ 
where $Q$ is
the PMT quantum efficiency and
$f$ the overall telescope optical factor. This factor includes
the optical filter transmission,  the corrector ring lens transmission (for telescopes
with such a lens), the mirror reflectivity, the mercedes (light collector) efficiency and
the camera shadow factor.

\noindent 4. {\it Conversion from sky background variance to photoelectrons:}
$$ n_{phe} = \frac{\sigma^2_{phe}}{1+ V_G}  $$
where $V_G$ is the PMT gain variance factor. $V_G$, which describes the
non poissonian effects induced by the PMT gain chain, is derived in PMT testing 
%\cite{pmt}
operations and from the manufacturer data sheets. It can be reasonably assumed equal
for all PMT's.

\noindent 5. {\it Conversion from photoelectrons to photon flux:}
$$ \Phi_{\gamma} = \frac{n_{phe}}{Q \cdot f \cdot \mathcal{A} \cdot \Delta t}  \qquad , $$
$\mathcal{A}$ is the pixel aperture and $\Delta t$ the sampling time 
slot (100\,ns). 
During this period only telescopes 3 and 4 in both eyes were equipped with a
corrector ring. The optical factors are 0.50 for these telescopes and 0.47 for the others. 
The average pixel apertures are 7.7 and 4.6 m$^2$ deg$^2$ for the telescopes with and without
corrector rings, respectively.

The M-UG6 filter at the aperture\cite{EA} bounds our measurement to the wavelength range of about
300-400 nm. The use of calibration constants provided by the drum results in a flux expressed in
terms of 370 nm-equivalent photons. 

\section{Results}
\label{sec:res}

Table \ref{allfluxes} gives the overall results of our analysis.
The high rate in telescope 6 at Los Leones arises because of artificial light from the town
of Malarg\"ue that is in its field of view. 
For this reason this telescope has been excluded from further analysis. Telescope 5 at 
Los Leones and telescope 2 at Coihueco are also affected by this source, but to a lesser extent.

The dispersions in table \ref{allfluxes} are calculated over all pixels as well as 
over the whole period. These dispersions result from two different contributions:
a) local variations depending on the photon flux seen by the individual pixel;
b) time variations induced by different global atmosphere transmission from night to night.

\begin{figure}[t]
\vskip -0.15cm
\centering
\begin{tabular}{p{.45\textwidth} p{.45\textwidth}}
\includegraphics[width=.45\textwidth]{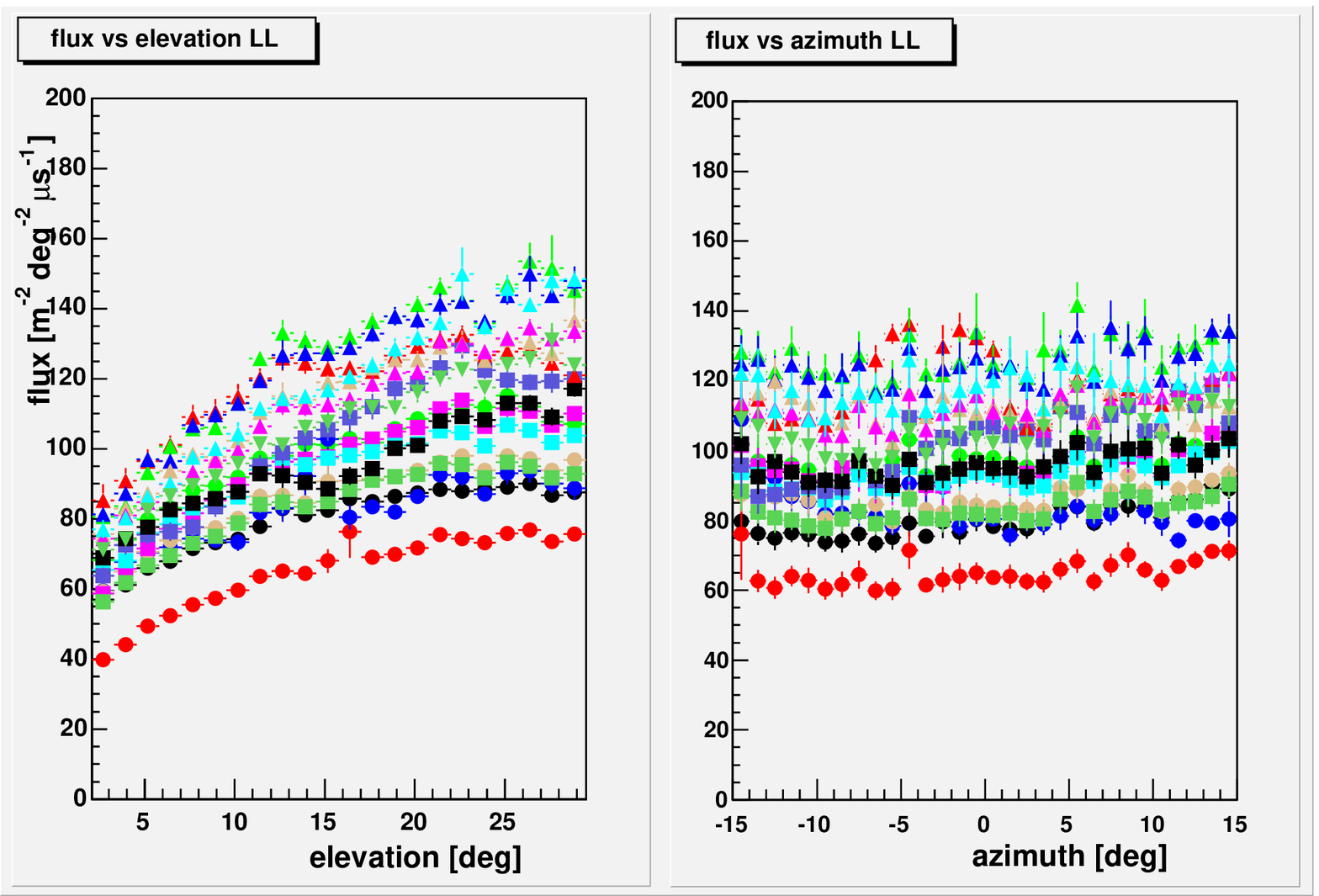} 
\caption{\label{elev_azi}{\small Sky background flux 
as a function of elevation (left) and azimuth (right) angles  for telescope 4 at Los Leones. 
Each marker refers to an individual night during the analysis period.}}
&
\includegraphics[width=.45\textwidth]{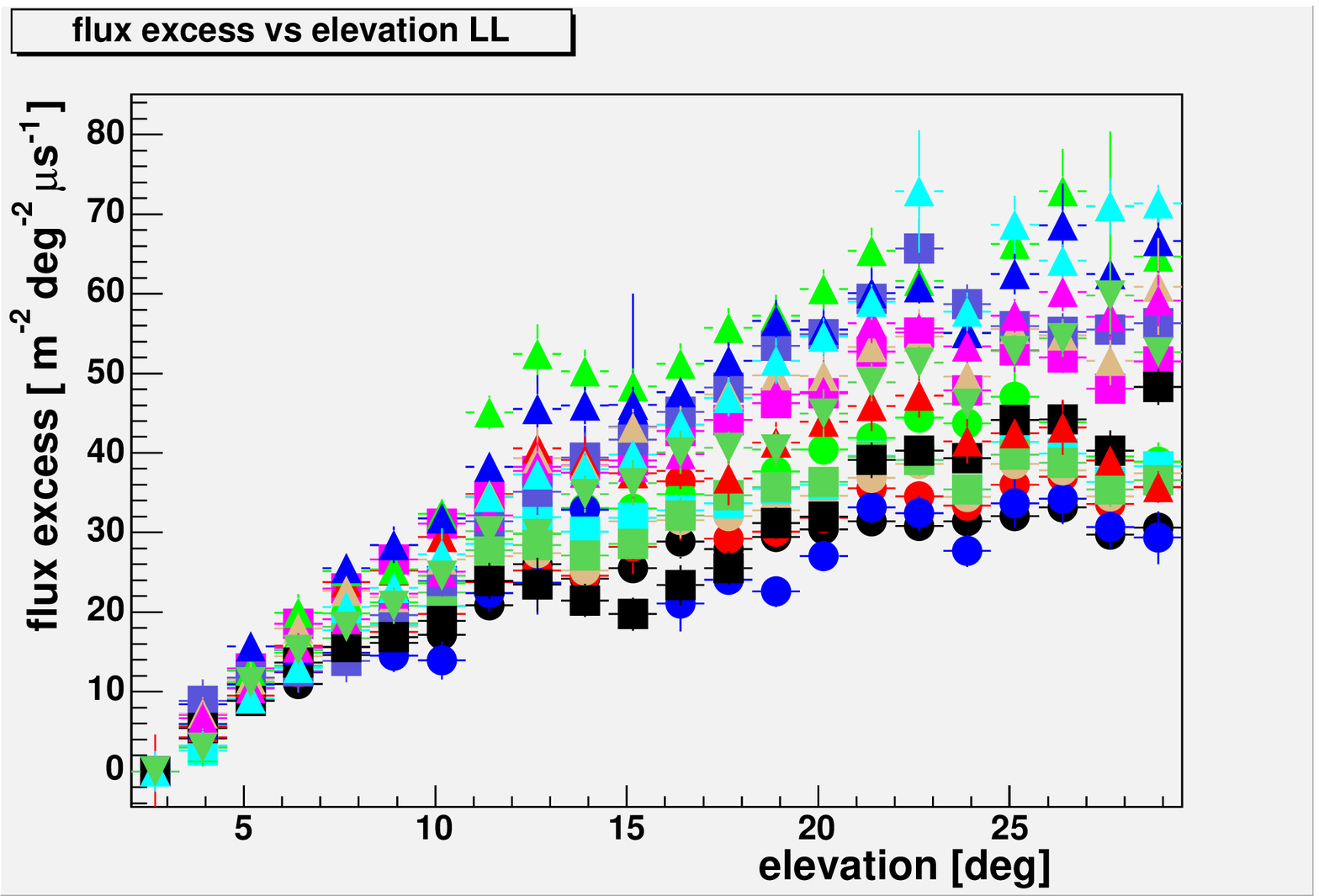}
\caption{\label{fex}
{\small Sky background flux excess with respect to the horizon,
as a function of elevation for telescope 4 at Los Leones.}}
\end{tabular}
\end{figure}

Local background variations can be studied by plotting the photon flux 
for each pixel. 
The most evident feature emerging from this analysis is a systematic up-down effect in the camera 
illumination.
%\begin{center}
\begin{table}[!ht]
\vskip -0.15cm
\begin{tabular}{||c||p{2.5cm}|p{2.5cm}||p{2.5cm}|p{2.5cm}||} 
\hline 
\multicolumn{1}{||c||}{} & 
\multicolumn{2}{|c||}{Flux at Los Leones (m$^{-2}$ deg$^{-2}$ $\mu$s$^{-1}$)} &
\multicolumn{2}{|c||}{Flux at Coihueco (m$^{-2}$ deg$^{-2}$ $\mu$s$^{-1}$)} \\
\hline 
\hline 
Telescope  & \makebox[2.5cm][c]{mean} & \makebox[2.5cm][c]{dispersion}  &  
\makebox[2.5cm][c]{mean} & \makebox[2.5cm][c]{dispersion} \\ \hline
\makebox[2.5cm][c]{1} & \makebox[2.5cm][c]{92}   & \makebox[2.5cm][c]{19}  &  
\makebox[2.5cm][c]{94} & \makebox[2.5cm][c]{15}   \\ \hline
\makebox[2.5cm][c]{2} &  \makebox[2.5cm][c]{93}   & \makebox[2.5cm][c]{19} &
\makebox[2.5cm][c]{97}  & \makebox[2.5cm][c]{28}    \\ \hline  
\makebox[2.5cm][c]{3} &  \makebox[2.5cm][c]{83}  & \makebox[2.5cm][c]{17} &  
\makebox[2.5cm][c]{80}  & \makebox[2.5cm][c]{24}   \\ \hline
\makebox[2.5cm][c]{4} &  \makebox[2.5cm][c]{85}  & \makebox[2.5cm][c]{17}   &  
\makebox[2.5cm][c]{77}  & \makebox[2.5cm][c]{20}  \\ \hline 
\makebox[2.5cm][c]{5} &\makebox[2.5cm][c]{101}  & \makebox[2.5cm][c]{17} & 
\makebox[2.5cm][c]{88}  & \makebox[2.5cm][c]{25}    \\ \hline 
\makebox[2.5cm][c]{6} & \makebox[2.5cm][c]{134}  & \makebox[2.5cm][c]{18} &  
\makebox[2.5cm][c]{91}  & \makebox[2.5cm][c]{27}   \\ \hline 
\end{tabular}
\caption{\small{Background fluxes in the two eyes. The values are averaged over the whole period and
over pixels in the camera. 
 }}
\label{allfluxes} 
\end{table}
This effect is shown in figure \ref{elev_azi} where the
sky background flux is shown as a function of elevation and azimuth
angles respectively for Telescope 4 at Los Leones. Each marker refers
to an individual run during the analysis period.  A similar behavior is found
for each telescope in both eyes. These graphs show that:

\noindent 1. the azimuth dependence is rather small;

\noindent 2. the elevation dependence is quite evident, showing an increase as elevation goes from 0$^\circ$
to 30$^\circ$;

\noindent 3. the two behaviors are roughly the same from run to run apart from a vertical offset. We assume that this 
offset is mainly determined by the atmospheric conditions of each night.
%***********************************************************************************
%***********************************************************************************

We have chosen the photon flux averaged over the first row of pixels in a camera as a measure of this offset. 
The corresponding elevation
is about 2$^\circ$; for this reason this offset will be referred to as the {\it horizontal flux}. 
Note that this offset is calculated for each run. Therefore the
corresponding flux with this offset subtracted gives the {\it flux excess} with respect to
the horizon.

In Figure \ref{fex} this flux excess is shown as a function of elevation for Telescope 4 at Los 
Leones. One can conclude
that, once the horizontal flux is subtracted, each night shows roughly the same behavior. 
Therefore the local component of the flux dispersion 
can be mainly attributed to a physical effect , i.e. the elevation dependence of the flux.

\begin{figure}[t]
\centering
\includegraphics[width=9cm]{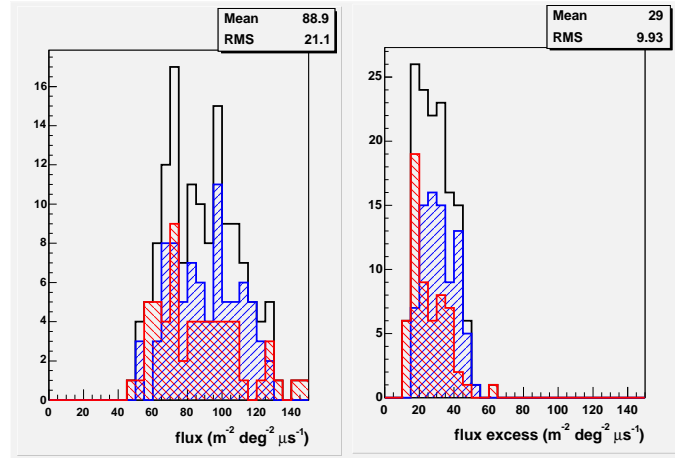}
\caption{\small{Sky background flux (left) and flux excess (right) distributions.
Each entry contains the relevant flux at each telescope averaged over all pixels for each night. The
dashed histograms refer to Los Leones (blue) and Coihueco (red), the open histogram to both sites.
Mirror 6 of Los Leones is excluded from this plot.}}
\label{fldistr}
\end{figure}

We also expect that the average of 
the flux excess over all pixels should exhibit small fluctuations from night to night.
To show this, we have calculated for each night the sky background flux
and the corresponding flux excess averaged over all pixels. In figure \ref{fldistr} 
these results are shown. The dispersions are 21 and 10 
photons / (m$^2$ deg$^2$ $\mu$s) for the total
flux and the flux excess respectively. Thus the dispersion of the flux excess is lower by
a factor of two. The difference between the two dispersions (in quadrature)
is an estimate of the dispersion due to the variation of atmospheric conditions
during the four month period considered in this analysis.  This component turns out to be
18 photons / (m$^2$ deg$^2$ $\mu$s), i.e. about 90\% of the total flux dispersion.

%***********************************************************************************
%***********************************************************************************

\section{Concluding remarks}
\label{sec:conc}

We have shown that a variance analysis allows the determination of the sky photon background flux from
FD data. The statistics from each night of data taking are enough to uncover local dependencies
of the photon flux. In particular, the change of flux with elevation angle is rather clear and
reproducible. Further studies are in progress to correlate this dependence with atmospheric properties 
(e.g. the horizontal aerosol attenuation). Correlations would be quite interesting as this would allow
us to deduce atmospheric parameters directly from FD data, thus allowing cross checks of independent
methods based on different monitoring devices.

This method,
if included in the standard post-run FD data processing, can allow for accurate monitoring
of the performance of all telescopes and possibly an estimate of atmospheric parameters 
directly associated with each FD run.


\begin{thebibliography}{9} 
\bibitem{EA}  J. Abraham {\it et al.}, The Auger Collaboration, Nucl. Instr. Meth. A523 (2004) 50-95 
\bibitem{HG} H. Gemmeke {\it et al.}, Proceedings of the 28th ICRC, Tsukuba, 2003, p. 891.
\bibitem{jeff} P. Bauleo {\it et al.}, contribution in these proceedings.

\end{thebibliography}
\end{document}